\documentclass[twocolumn,floatfix]{aastex62}

\usepackage{savesym}
\savesymbol{tablenum}
\usepackage{siunitx}
\usepackage{bm}
\usepackage{mathtools}

\graphicspath{{./}{figures/}}

\submitjournal{ApJL}
\begin{document}

\title{Ram-pressure Stripping of a Kicked Hill Sphere:
Prompt Electromagnetic Emission \\ from the Merger of Stellar Mass Black Holes in an AGN Accretion Disk}

\correspondingauthor{Barry McKernan}
\email{bmckernan@amnh.org}

\author{B. McKernan}
\affiliation{Department of Astrophysics, American Museum of Natural
  History, Central Park West at 79th Street, New York, NY 10024}
\affiliation{Department of Science, Borough of Manhattan Community
  College, City University of New York, New York, NY 10007}
\affiliation{Physics Program, The Graduate Center, City University of New York, New York, NY 10016}

\author{K.E.S. Ford}
\affiliation{Department of Astrophysics, American Museum of Natural
  History, Central Park West at 79th Street, New York, NY 10024}
\affiliation{Department of Science, Borough of Manhattan Community
  College, City University of New York, New York, NY 10007}
\affiliation{Physics Program, The Graduate Center, City University of New York, New York, NY 10016}

\author{I. Bartos}
\affiliation{Department of Physics, University of Florida, Gainesville, FL 32611}

\author{M.J. Graham}
\affiliation{Cahill Center for Astronomy \& Astrophysics, California Institute of Technology, 1200 E California Blvd, Pasadena, CA 91125}

\author{W. Lyra}
\affiliation{Department of Physics and Astronomy, California State University Northridge, 18111 Nordhoff Street, Northridge, CA 91330}
\affiliation{Jet Propulsion Laboratory, California Institute of Technology, 4800 Oak Grove Drive, Pasadena, CA 91109}

\author{S. Marka}
\affiliation{Columbia Astrophysics Laboratory, Columbia University, New York, NY 10027}

\author{Z. Marka}
\affiliation{Columbia Astrophysics Laboratory, Columbia University, New York, NY 10027}

\author{N.P. Ross}
\affiliation{Institute for Astronomy, University of Edinburgh, Royal Observatory, Blackford Hill, Edinburgh EH9 3HJ, UK}

\author{D. Stern}
\affiliation{Jet Propulsion Laboratory, California Institute of Technology, 4800 Oak Grove Drive, Pasadena, CA 91109}

\author{Y. Yang}
\affiliation{Department of Physics, University of Florida, Gainesville, FL 32611}



\begin{abstract}

    Accretion disks around supermassive black holes (SMBH) are promising sites for stellar mass black hole (BH) mergers due to mass segregation and merger acceleration by disk gas torques. Here we show that a  GW-kick at BH merger causes ram-pressure stripping of gas within the BH Hill sphere. If $R_{H}\geq H$, the disk height, an off-center UV flare at $a_{\rm BH} \sim 10^{3}r_{g}$ emerges within $t_{\rm UV} \sim \rm{O}(2 {\rm days})(a_{\rm BH}/10^{3}r_{g})(M_{\rm SMBH}/10^{8}M_{\odot})(v_{\rm kick}/10^{2}\rm{km/s})$ post-merger and lasts O$(R_{H}/v_{\rm kick}) \sim \rm{O}(5 t_{\rm UV}$). The flare emerges with luminosity O($10^{42}{\rm erg/s})(t_{\rm UV}/2{\rm days})^{-1}(M_{\rm Hill}/1M_{\odot})(v_{\rm kick}/10^{2}{\rm km/s})^{2}$. AGN optical/UV photometry alters and asymmetric broad emission line profiles can develop after weeks. If $R_{H}<H$, detectability depends on disk optical depth. Follow-up by large optical sky surveys is optimized for small GW error volumes and for LIGO/Virgo triggers $>50M_{\odot}$.

\end{abstract}


\section{Introduction}
Advanced LIGO \citep{aLIGO} and Advanced Virgo \citep{adVirgo} are revealing a surprisingly numerous population of merging stellar mass black holes (BHs; \citealt{LIGO18}), including a previously undetected population of black holes with masses $>20 M_\odot$. In the local Universe, BH density seems greatest in our own Galactic nucleus \citep{Hailey18,Generozov18}, consistent with previous conjectures \citep{Morris93,Miralda00}. Thus, a promising channel for the LIGO/Virgo BH mergers are active galactic nucleus (AGN) disks. This is because a fraction of the orbiting nuclear population (including BHs), are geometrically coincident with the AGN disk and another fraction is ground-down into the disk \citep{Syer91,Arty93,GT04}. Torques from AGN disk gas drive binary formation, migration, and mergers in these populations \citep{McK12,McK14,Bellovary16,Bartos17,Stone17,McK17,Secunda19,McK19}. Here we show there can be a  prompt, bright UV/optical counterpart from AGN disks after a kicked BH merger. Follow-up by wide-area optical photometric surveys (e.g. ZTF) is optimized for LIGO/Virgo triggers $>50M_{\odot}$ with small error volumes. Detection of such signatures would assign specific galactic counterparts to GW sources and probe AGN disk interior conditions for the first time. 

\section{Hill sphere reaction post-merger}
Mass is lost in a BH merger during chirp and ringdown as gravitational waves (GWs) carry away energy and angular momentum. The final mass ${M_{f}}$ is \citep{Tichy08}
\begin{equation}
M_{\rm f}=M_{b} \left[ 1-0.2 \nu -0.208\nu^{2}(a_{1}+a_{2})\right]
  \end{equation}
where $\nu \equiv \mu/M_{b}=q_{b}/(1+q_{b})^{2}$ is the symmetric mass ratio of binary $M_{b} \equiv M_{1}+M_{2}$ where $q_{b} \equiv M_{2}/M_{1}$, and $a_{1,2}$ are the spin magnitudes of masses $M_{1,2}$. Typically $\Delta M \equiv (M_b - M_{\rm f})/M_b \sim 0.05$ for $q_{b} \sim 1$ and small $a_{1},a_{2}$. The sphere of influence of a binary BH system of mass $M_{b}$ in orbit around a super-massive black hole of mass $M_{\rm SMBH}$ is given by the Hill radius ($R_{H} \equiv r(q/3)^{1/3}$), where $r$ is the semi-major axis of the binary center-of-mass around the SMBH and $q \equiv M_{b}/M_{\rm SMBH}$. Post-merger $R_{H}$ decreases at light speed by
\begin{eqnarray}
\Delta R_{H} &\approx& r \left(\frac{M_{b}}{3M_{\rm
      SMBH}}\right)^{1/3} \left(\frac{\Delta M}{3M_{b}}\right) \approx R_{H} \left(\frac{\Delta M}{3M_{b}} \right)
  \end{eqnarray}
for small $\Delta M$. 
Several effects result. First, gas within the Hill sphere that is orbiting too fast for new mass $M_{f}$ moves outward, self-colliding. Second, gas formerly inside $R_{H}$ now collides with gas orbiting the SMBH. Third, the post-merger BH accretes the low angular-momentum component of self-shocked Hill sphere gas in a burst. A jet or beamed outflow adds luminosity $\eta \dot{M}_{f} c^{2}$ to the emerging, observable hot-spot outlined below, where $\eta$ is the accretion efficiency onto $M_{f}$.\\ 
Separately, a GW-kick at merger \citep[e.g.][]{Baker08,Zlochower11} causes ram-pressure stripping of the original Hill sphere gas as it collides with a comparable mass of disk gas. Many of these effects have been studied in the context of SMBH mergers in gas disks  \citep[e.g.][]{Bog08,Kocsis08,Shields08,Megevand09,Oneill09,Bode10}. However the physics of the kicked Hill sphere colliding with surrounding gas has no direct analogy in circumbinary disks for SMBH mergers.

\subsection{Collisions involving Hill sphere gas}
Post-mass loss, parcels of gas within the Hill sphere are at the pericenter of a new, eccentric orbit. The Hill sphere gas will self-collide on timescales $>$O($t_{\rm orb}$) \citep{Oneill09}. Supersonic collisions occur deepest in the Hill sphere ($r=R_{1}<R_{H}$), where the collisional Mach number (${\cal M}$) is
\begin{equation}
{\cal{M}} \approx \frac{1}{2} \frac{1}{(r/R_{H})^{1/2}}\frac{q^{1/3}}{r(r_{g})^{1/2}}\left(\frac{c}{c_{s}}\right)\left(\frac{\Delta M}{M_{b}}\right)
\end{equation}
where $c_{s}$ is the gas sound speed and $\Delta v/v_{\rm K,Hill} \approx \Delta M/2M_{b}$ where $v_{\rm K,Hill}(r) =c/\sqrt{r_{g,b}}$ is the gas Keplerian velocity with $r_{g,b}=GM_{b}/c^{2}$. For $q \geq 10^{-4}$, supersonic collisions can extend to $R_{1} \sim R_{H}$. If ${\cal M}\gg 1$, ($r \ll R_{!}$) the post-shock temperature is $T \sim (5/16){\cal M}^{2}T_{\rm disk}$ from Rankine-Hugoniot jump conditions, where $T_{\rm disk}$ is the model disk temperature at the merger radius. If ${\cal M} \sim 1+\epsilon$, ($r \sim R_{1}$), then $T \sim (1+\epsilon)T_{\rm disk}$. The collisions at $R_{1}<R_{H}$ generate energy $E_{1}\approx (3/2)M_{\rm Hill}(R_{1}/R_{H})^{3}(k_{B}/m_{H})T_{\rm disk}$ on orbital timescales around $M_{f}$.\\
In the sub-sonic collision zone ($R_{1}<r<R_{H}-\Delta R_{H}$), ${\cal M} <1$ and the average gas temperature is $T \sim (1/3)(m_{H}/k_{B}) (\Delta M \Delta v_{2}/M_{b})^{2}$, where $\Delta v_{2}$ is the velocity difference between [$R_{1},R_{H}$]. Subsonic collisions contribute $E_{2} \approx (1/2)M_{\rm Hill}(\Delta M \Delta v_{2}/M_{b})^{2}$. Gas between $R_{H}-\Delta R_{H}$ and $R_{H}$, with total mass $\Delta M_{\rm Hill}$, collides with gas orbiting the SMBH at velocity differential $\Delta v(r) \approx (c/2r(r_{g})^{1/2})(q/3)^{1/3}$, yielding a average uniform temperature of $T=(1/3)(m_{H}/k_{B})(\Delta v(r))^{2}$ and energy $E_{3} \approx (1/2)\Delta M_{\rm Hill}\Delta v(r)^{2}$.\\
The luminosity of the self-collisions above are limited by the timescales on which the gas collides, which could be $\gg t_{\rm orb}$ \citep{Oneill09}. The most luminous EM contribution post-merger occurs as the kicked BH remnant attempts to carry its original Hill sphere gas with it. This is because: 1) it involves most of $M_{\rm Hill}$ and 2) the timescale of the disk response is short ($R_{H}/v_{\rm kick}$) compared to most self-collision timescales. Physically, once the Hill sphere gas  collides with an equivalent mass of disk gas (in time $R_{H}/v_{\rm kick}$), much of the original Hill sphere gas is lost via ram pressure stripping. As a result, energy $E_{\rm kick}=1/2M_{\rm H} v_{kick}^{2} \sim 10^{47}{\rm erg} (M_{\rm Hill}/1M_{\odot})(v_{\rm kick}/10^{2}{\rm km/s})^{2}$ is dissipated via shocks at temperature $T \sim O(10^{5}){\rm K} (v_{\rm kick}/10^{2}{\rm km/s})^{2}$ over a duration O($R_{H}/v_{\rm kick}$). Here we assume an adiabatic shock; however, the timescales for large mass SMBH, especially for merging binaries on large orbits, imply non-adiabatic processes will be important. For such circumstances, our estimates represent upper limits.

\subsection{Disk opacity}
Photons liberated by gas shock heating can be scattered inside the disk before escape. If the merger occurs where gas pressure dominates then the disk atmospheric density is $\rho(z)=\rho_{0}\exp{(-z^{2}/2H^{2})}$ where $\rho_{0}$ is the mid-plane density, $z$ is the vertical distance, and $H$ is the disk height. The mean free path length of photons in the disk is $\ell=1/(\kappa \rho)$, where $\kappa$ is the disk opacity. Total path length $L$ travelled by photons to the disk surface is $L=N \ell$ where $N=H^{2}/\ell^{2}$ is the average number of steps. Therefore, $L=H^{2}/\ell$ and the mean photon travel time is $t_{\rm Hill}=L/c$. Assuming constant dissipation per unit optical depth, the disk surface temperature ($T_{\rm eff}$) is $T_{\rm eff} \approx ((3/8)\tau + 1/(4\tau))^{-1/4}T_{0}$ where $\tau=\kappa \Sigma/2$ and $T_{0}$ is the midplane temperature \citep{SG03}. Now we consider two illustrative AGN disk models where $R_{H} \geq H$ and $R_{H} < H$. This allows us to quantify EM signatures at the order-of-magnitude level.

\subsection{$R_{H} \geq H$: prompt UV flare}
\label{sec:lessthanh}
Consider a kicked $M_{b}=65\, M_{\odot}$ binary located at $r=10^{3}r_{g}$ from a $M_{\rm SMBH}=10^{9}\, M_{\odot}$ SMBH, so  $R_{H} \sim 3r_{g}$. For a \citet{Thompson05} disk model at radius $r = 10^{3}r_{g}$, $H/r \sim 10^{-3}$, so $H \sim r_{g}$ and $R_{H}>H$ in this example. The volume of gas in the Hill sphere is 
\begin{equation}
    V_{\rm gas}=\frac{4}{3}\pi R_{H}^{3} -\frac{2}{3}\pi(R_{H}-H)^{2}[3R_{H}-(R_{H}-H)].
\end{equation}
Here, $V_{\rm gas}/V_{\rm Hill} \sim 0.5$, so $M_{\rm Hill}=V_{\rm gas}\rho \sim 0.8M_{\odot}$.

The ram-pressure stripping of the kicked Hill sphere gas releases shock energy $E_{\rm kick}=1/2M_{\rm Hill} v_{\rm kick}^{2} \sim 10^{47}\rm{erg} (M_{\rm Hill}/1M_{\odot}) (v_{\rm kick}/10^{2}{\rm km/s}) $ at $T \sim (m_{H}/k_{B})v_{\rm kick}^{2} \sim O(10^{5})(v_{\rm kick}/10^{2}{\rm km/s})$K over timescale $O(R_{H}/v_{\rm kick}) \sim 6\rm{mo}$, yielding a UV luminosity $\sim 10^{41}{\rm erg/s}$. If disks similar to the \citet{Thompson05} model are found around smaller mass SMBH, the timescale $R_{H}/v_{\rm kick}$ drops considerably and the luminosity can reach $\sim 10^{42}\rm{erg/s}$ around $M_{\rm SMBH} \sim 10^{6}M_{\odot}$ (see below).

\subsection{$R_{H}<H$: delayed, weak flare}
A $M_{b}=65\, M_{\odot}$ binary located at $r=10^{3}\, r_{g}$ from a $M_{\rm SMBH}=10^{8}\, M_{\odot}$ SMBH has a Hill sphere of radius $R_{H} \sim 6r_{g}$. In a \citet{SG03} disk model at $r = 10^{3}\, r_{g}$,  $H/r \sim 10^{-2}$ so $H \sim 10\, r_{g}$ and $R_{H}<H$. Therefore $V_{\rm gas} = V_{\rm Hill}$ and $M_{\rm Hill}=V_{\rm Hill}\, \rho \sim 1.5M_{\odot}$. The photon diffusion length in the disk is $\langle L \rangle= \langle H \rangle^{2}/\ell$ where $\langle H \rangle$ spans $[H-R_{\rm H},H]$, or $[2.1,10]r_{g}$ in this example. For an exponential atmosphere with $z\sim H$, $\langle L \rangle=\langle H \rangle^{2}/\ell \sim [0.4,22] \times 10^{17}\, {\rm cm}$ and the photon travel time spans $t_{\rm Hill}=\langle L \rangle /c$ or $t_{\rm Hill} \sim [15\, {\rm day}, 2.3\, {\rm yr}]$. For $\tau_{0} \sim 10^{4}$, $T_{\rm eff} \sim 0.1 T_{0}$. So, while UV/optical photons emerge from the kick-shock on a timescale O(month) post-merger, the reprocessed optical signature is smeared out over several months with luminosity $\leq 10^{41}\, {\rm erg\, s^{-1}}$. Upper limits on follow-up of LIGO/Virgo search volumes therefore constrains ($H,\rho$) in AGN disks.

\subsection{New thermal emission from kicked hot spot}
The temperature of an unperturbed thin AGN disk is $T(r)=T_{\rm max} r^{-3/4}$, where $T_{\rm max}\sim 6 \times 10^{5}\, (M_{\rm SMBH} / 10^{8}\, M_{\odot})^{1/4}\ (\dot{m}/\dot{M}_{\rm Edd})^{1/4}$K and $\dot{M}_{\rm Edd}$ is the Eddington accretion rate. For $R_{H}>H$, $T_{\rm eff}$ can be $>T_{\rm max}$ if $\tau_{0}$ is small. The emitting area of the kicked Hill sphere is $(R_{H}/R_{\rm disk})^{2} \sim 5 \times 10^{-7}$ smaller than the disk (if $R_{\rm disk} \sim 10^{4}\, r_{g}$). The kicked hot spot has a temperature dependent only on $v_{\rm kick}^{2}$. Thus, a kicked merger in a AGN disk around a small mass SMBH can generate a short-lived luminous hot spot which could dominate continuum emission in the UV or optical bands.

\begin{figure}
\begin{center}
\includegraphics[width=8.0cm,angle=0]{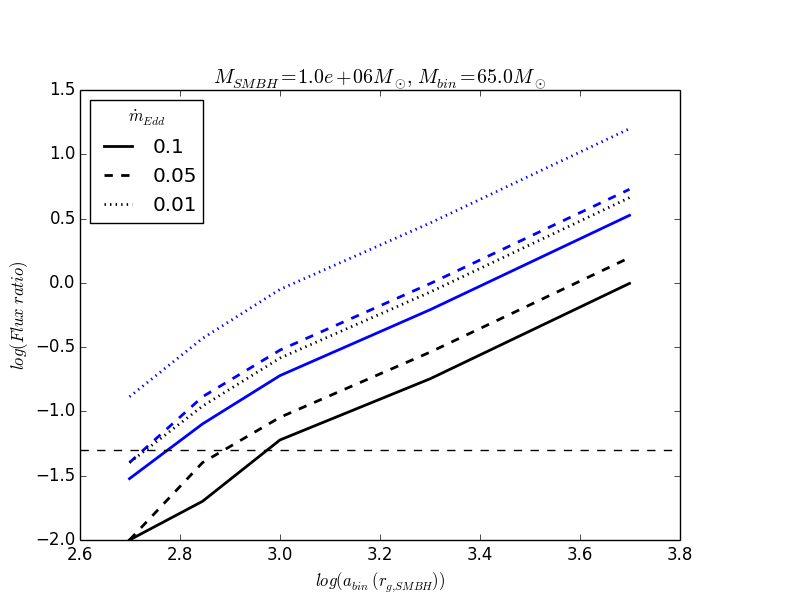}
\end{center}
    \caption{Ratio of optical (ZTF $g$-band) flux (black curves) and NUV ({\it GALEX} band) flux (blue curves) from kicked Hill sphere hot spot to (ZTF $g$-band, {\it GALEX} NUV) flux from a unperturbed AGN disk around a $M_{\rm SMBH}=10^{6}\, M_{\odot}$ SMBH, as a function of binary distance ($a_{\rm bin}$) from SMBH. We assume $M_{\rm b}=65\, M_{\odot}$,$\Delta M \sim 0.05$ \& $v_{\rm kick}=10^{2}$km/s. Horizontal dashed line corresponds to a flux increase of $\sim 5\%$. Curves correspond to $M_{\rm SMBH}$ accreting at fractions $0.1$(solid), $0.05$(dashed), $0.01$(dotted) of the Eddington rate. At $a_{b} \geq 10^{3}r_{g}$, the change in optical/NUV flux is detectable for $R_{H}>H$.
    \label{fig:1e6}}
    
\end{figure}

\begin{figure}
\begin{center}
	\includegraphics[width=8.0cm,angle=0]{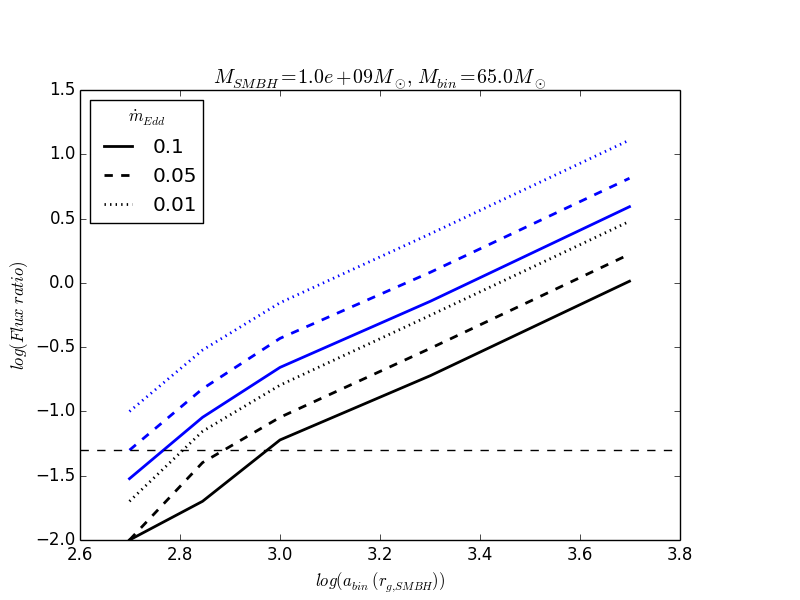}
\end{center}
    \caption{As Fig.~\ref{fig:1e6} except for $M_{\rm SMBH}=10^{9}M_{\odot}$.}
    \label{fig:1e9}
\end{figure}

Fig.~\ref{fig:1e6} shows (black curves) the ratio of rest-frame $g$-band optical flux from the hot-spot to that emitted by the AGN disk and (blue curves) the corresponding {\it GALEX} NUV flux ratio. The curves assume a kicked $M_{b}=65\, M_{\odot}$ BH merger ($v_{\rm kick} \sim 10^{2}\rm{km/s}$) in a disk around a $M_{\rm SMBH}=10^{6}\, M_{\odot}$ SMBH accreting at $[0.01,0.1]\, \dot{M}_{\rm Edd}$, with $R_{H}>H$ so flux escapes. Fig.~\ref{fig:1e9} is as Fig.~\ref{fig:1e6}, but for $M_{\rm SMBH}=10^{9}M_{\odot}$. Horizontal dashed lines show the detectability threshold of a $5\%$ flux change. Figs.~(\ref{fig:1e6}) and (\ref{fig:1e9}) show the  post-merger Hill sphere kick can be a substantial fraction of (or even dominate) $g$-band/NUV emission from the AGN disk if $a_{b} \geq 10^{3}r_{g}$. ZTF and LSST can easily detect such changes ($>5\%$) in $g$-band for well-detected sources. If $R_{H}<H$, the hot-spot is obscured. The curves in Figs.~\ref{fig:1e6} and \ref{fig:1e9} drop by a factor $O(T_{0}/T_{\rm eff})$ or $-1.2$ in log flux ratio for $T_{0}/T_{\rm eff} \sim 2$. Mergers in lower accretion rate AGN disks with $R_{H}<H$ may be detectable for small $T_{0}/T_{\rm eff}$ if $\dot{M}_{\rm Edd}=0.01$. 

\subsubsection{Asymmetry in BLR line profiles}
The broad line region (BLR) in AGN is modelled as a distribution of clumpy clouds at median radius $R_{BLR}$ \citep{Krolik81,Netzer10}. The hot-spot on the disk photosphere emerges in days ($R_{H}>H$), whereupon the BLR is asymmetrically illuminated in time $R_{BLR}/c$ (a few weeks). Off-center illumination generates a uniform asymmetry in all broad line components, unlike central AGN variability. If the hot spot persists, the asymmetric illumination washes over the BLR in several months.  

\section{Strategy for follow-up of a LIGO/Virgo GW detection volume}

LIGO--Virgo are detecting BH mergers at a rate 1/wk in the third observing run (O3)\footnote{https://gracedb.ligo.org/latest/}. If BH mergers are preferentially associated with AGN, we must optimize searches for the EM signatures above.  If LIGO/Virgo releases binary mass estimates ($>50M_{\odot}$) with GW triggers, we increase the likelihood that $R_{H}>H$ so an EM counterpart is detectable.
 
Optimal search involves rapid wide-field UV/optical follow-up of small LIGO/Virgo error boxes when $M_{b}> 50M_{\odot}$.  UV is preferred since the fractional signature is more pronounced (Figs. 1 and 2), though such capabilities are currently lacking. Candidates displaying photometric jumps may have $R_{H}>H$ and optical spectroscopic follow-up can search for broad line asymmetries, indicating an off-center hot-spot. If $R_{H}<H$ then the disk hot-spot shows up weeks/months  post-GW trigger. Optical surveys can detect photometric changes if $R_{H}>H$ and $a_{b}>2000\, (4000)\, r_{g}$ in AGN disks accreting at $0.01\, (0.1)\, \dot{M}_{\rm Edd}$. 
 
A vetted AGN catalog is a good starting point for optical/UV follow-up. \citet{Assef18} provides a catalog consisting of 4.5 million AGN candidates across the full extragalactic sky ($\approx$150 candidates deg$^{-2}$) with 90\% reliability (their ``R90'' catalog), as well as a lower reliability catalog of 21 million AGN candidates across the full extragalactic sky ($\approx$700 candidates deg$^{-2}$) with 75\% completeness; their ``C75'' catalog). Both catalogs derive from the AllWISE Data Release \citep{Wright10}, and are currently the widest area published lists of AGN candidates across the full sky. 
  
\section{Conclusions}
A merging BH binary in an AGN disk generates a prompt set of EM signatures if the Hill sphere radius is greater than the AGN disk height ($R_H>H$). The GW-kick causes ram pressure stripping of the Hill sphere. The resulting shock dominates the EM response \& has no analog in SMBH mergers. Searches, including non-detections, constrain ($H,\rho$) in AGN disks. If $R_{H}<H$, detectability depends heavily on disk opacity. UV searches are optimal, but small LIGO/Virgo error boxes can be efficiently searched by large optical surveys with photo-z selected AGN catalogs. LIGO/Virgo triggers should include 'large mass' ($M_{b}>50M_{\odot}$) estimates to optimize EM follow-up.

\section{Acknowledgements}
BM dedicates this paper to the memory of his mother, Treasa McKernan. BM \& KESF are supported by NSF 1831412. The work of DS was carried out at the Jet Propulsion Laboratory, California Institute of Technology, under a contract with NASA. Thanks to the referee for highlighting the GW kick \& participants in the 1st workshop on stellar mass BH mergers in AGN disks, held March 11-13, 2019 and sponsored by the CCA at the Flatiron Institute in New York City.

\end{document}